\documentstyle[epsfig]{l-aa}

\newcommand\yr{$\rm yr^{-1}$}
\def\ltsima{$\; \buildrel < \over \sim \;$}
\def\simlt{\lower.5ex\hbox{\ltsima}}
\def\gtsima{$\; \buildrel > \over \sim \;$}
\def\simgt{\lower.5ex\hbox{\gtsima}}

\begin{document}
   
   \thesaurus{13.25.2 -- 11.03.4}

   \title{Spatially resolved spectroscopy of the
          cooling flow
          cluster PKS 0745$-$191 with BeppoSAX}

   \author{Sabrina De Grandi
          \inst{1}
   \and Silvano Molendi
	  \inst{2}
}

   \offprints{degrandi@merate.mi.astro.it}

   \institute{Osservatorio Astronomico di Brera, via Bianchi 46,
        I-23807 Merate (LC), Italy
   \and Istituto di Fisica Cosmica, CNR, via Bassini 15, 
       I-20133 Milano, Italy
}

   \date{Received / Accepted }

   \maketitle

\markboth{S. De Grandi and S. Molendi}{BeppoSAX observation of PKS0745}

\begin{abstract}
We present results from a BeppoSAX observation of the cooling flow
cluster PKS 0745$-$191 (z$=$0.1028).  By performing spatially resolved
spectroscopy, we find that the projected temperature profile is
consistent with being constant.  We can rule out, at more than the
99$\%$ confidence level, a temperature decrement of a factor 2 when
going from the cluster core out to 1.2 Mpc.  On the contrary, the
projected metal abundance is found to drop from 0.4 (solar units)
within the cluster core to 0.2 (solar units) at radii larger than 300
kpc, this decrement is significant at more than the 99.9$\%$
confidence level.


\end{abstract}

\section {Introduction}

PKS 0745$-$191 (hereafter PKS0745) is an X-ray luminous, relatively
near (z$=$ 0.1028), cluster of galaxies.  The proximity of this cluster 
to the galactic plane ($\sim3^o$) has hampered studies in the optical 
band.  On the contrary much attention has been devoted to the X-ray 
emission of PKS0745.  Early observations (Fabian et al. 1985) showed 
that PKS0745 hosts a particularly massive cooling flow, 
$\dot M \sim 1000 M_\odot$\yr.
ROSAT PSPC and HRI images of PKS0745 (e.g. Allen, Fabian \& Kneib 1996
, hereafter A96) show that the cluster has an elliptical shape.  No
significant structure is seen on large scales, indicating that quite
likely PKS0745 is not undergoing a merger event.  David et al. (1993),
using Einstein MPC data, report a global temperature of
8.5$^{+1.2}_{-0.8}$ keV for PKS0745.  A96, by modeling ASCA spectra
accumulated in concentric annuli, find evidence of a temperature
increase with increasing distance from the cluster core.  By
performing a full multiphase analysis of the same data, A96 find that
the temperature of the ambient gas is consistent with having a flat
radial profile. Thus the hardening of the spectrum seen with
increasing radius is totally attributed to the different fraction of
the cooling-flow emission included in the spectra accumulated from
different annuli.  Unfortunately the ambient temperature profile
cannot be constrained tightly using ASCA data.  Allen \& Fabian
(1998), by performing a multiphase analysis, measure an average metal
abundance of 0.35$^{+0.04}_{-0.03}$ for PKS0745.  A96, again using
ASCA data, do not find compelling evidence of an abundance gradient in
PKS0745.

In this Letter we report a recent BeppoSAX observation of PKS0745.
We use our data to perform an independent and more precise measurement
of the temperature and metal abundance profiles of this cluster.  We also
present the first temperature and metal abundance maps of PKS0745.
The outline of the Letter is as follows.  In section 2 we give some
information on the BeppoSAX observation of PKS0745 and on the data
preparation.  In section 3 we present the analysis of the integrated
spectrum of PKS0745.  In section 4 we present spatially resolved
measurements of the temperature and metal abundance.  In section 5 we
discuss our results and compare them to previous findings.
Throughout this Letter we assume H$_{o}$=50 km s$^{-1}$Mpc$^{-1}$
and q$_{o}$=0.5.
 
\section {Observation and Data Preparation}
The cluster PKS0745 was observed by the BeppoSAX satellite (Boella et
al. 1997a) between the 23$^{rd}$ and the 25$^{th}$ of October 1998.
We will discuss here data from the MECS instrument only, the PDS
instrument (Frontera et al. 1997), working in the 13-200 keV band, did
not detect any significant emission from the cluster.  The MECS
(Boella et al. 1997b) is presently composed of two units (after the
failure of a third one), working in the 1--10 keV energy range. At
6~keV, the energy resolution is $\sim 8\%$ and the angular resolution
is $\sim$0.7$^{\prime}$ (FWHM).  Standard reduction procedures and
screening criteria have been adopted to produce linearized and
equalized event files.  Data preparation and linearization was
performed using the {\sc Saxdas} package.  The effective MECS exposure
time of the observation was 9.5$\times$10$^4$ s.  The observed
countrate for PKS0745 was 0.645$\pm$0.003 cts/s for the 2 MECS units.

All MECS spectra discussed in this Letter have been background
subtracted using spectra extracted from blank sky event files in the
same region of the detector as the source. The energy range considered
for spectral fitting is always 2-10 keV.  All spectral fits have been
performed using XSPEC Ver. 10.00.  Quoted confidence intervals are
$68\%$ for 1 interesting parameter (i.e. $\Delta \chi^2 =1$), unless
otherwise stated.

\section{Spatially averaged spectral analysis} 

We have extracted a MECS spectrum, from a circular region of
8$^{\prime}$ radius (1.2 Mpc), centered on the emission peak. From the
ROSAT PSPC radial profile, we estimate that about 90$\%$ of the total
cluster emission falls within this radius.  The spectrum has been
fitted with a one temperature thermal emission component plus a
cooling flow component (MEKAL and MKCFLOW codes in the XSPEC package),
absorbed by a galactic line of sight equivalent hydrogen column
density, $N_H$, of 4.24$\times 10^{21}$ cm$^{-2}$ (Dickey \& Lockman
1990).  All parameters of the cooling flow component, except for the
mass deposition rate were constrained: the minimum temperature was
fixed at 0.1 keV, the maximum temperature, and the metal abundance
were set to be equal to the temperature and the metal abundance of the
MEKAL component.  The model yields an acceptable fit to the data,
$\chi^2 =$ 162.7 for 167 d.o.f. The best fitting values for the
temperature and the metal abundance are respectively, 8.5$\pm$0.6 keV
and 0.38$\pm$0.03, solar units. Not surprisingly, given the adopted
spectral range (2-10 keV), the mass deposition rate is rather ill
constrained $\dot M = 760\pm 460 M_\odot$\yr.  This value is however
in agreement with values derived by deprojecting the ROSAT PSPC
surface brightness profile or by performing spectral fits to ROSAT
PSPC and ASCA data (see A96).

\begin{figure}
\vspace{-0.19in}
\centerline{
      \hbox{
      \psfig{figure=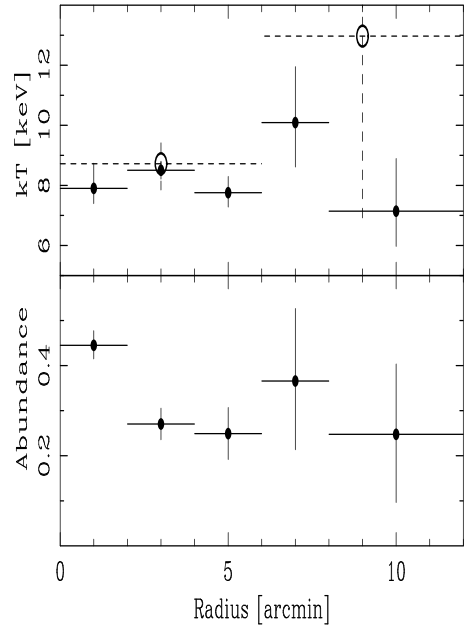,height=9cm,width=6.7cm,angle=-90}
}}
\vspace{-0.19in}
\caption{
{\bf Top Panel}: projected radial temperature profile.
The filled and open circles indicate respectively the
measurements obtained by fitting the continuum spectrum and 
from the energy of the centroid of the Fe K$_{\alpha}$ line. 
{\bf Bottom Panel}: projected radial abundance profile. 
}
\label{figmg7}
\end{figure}
\section{Spatially resolved spectral analysis} 
The spectral distortions introduced by the energy dependent MECS PSF,
when performing spatially resolved spectral analysis, have been taken
into account using the method described in Molendi et al. (1999),
hereafter M99, and references therein.

We have accumulated spectra from 5 annular regions centered on the
X-ray emission peak, with inner and outer radii of
0$^{\prime}$-2$^{\prime}$, 2$^{\prime}$-4$^{\prime}$,
4$^{\prime}$-6$^{\prime}$, 6$^{\prime}$-8$^{\prime}$ and
8$^{\prime}$-12$^{\prime}$.  A correction for the absorption caused by
the strongback supporting the detector window has been applied for the
8$^{\prime}$-12$^{\prime}$ annulus, where the annular part of the
strongback is contained. For the 6$^{\prime}$-8$^{\prime}$ region,
where the strongback covers only a fraction of the available area, we
have chosen to exclude the regions shadowed by the strongback.  We
have fitted each spectrum, except the one extracted from the innermost
region, with a MEKAL model absorbed by the galactic N$_H$, of
4.24$\times 10^{21}$ cm$^{-2}$.  In the spectrum from the
0$^{\prime}$-2$^{\prime}$ (0-0.3 Mpc) region we have included a
cooling flow component, the parameters of this component have all been
fixed except for the mass deposition rate, as in the fitting of the
integrated spectrum (see section 3).  The temperature and abundance we
derive for the innermost region are respectively 7.9$^{+0.8}_{-0.5}$
keV and 0.45 $\pm$ 0.03, solar units, the mass deposition rate is
found to be $\dot M = 680^{+430}_{-350} M_\odot$\yr.  If we fix the
mass deposition rate to the value of $1000 M_\odot$\yr, estimated by
A96, we obtain somewhat tighter constrains on the temperature
8.5$\pm$0.3 keV, while the abundance estimate 0.46$\pm$ 0.03, solar
units, is practically unaffected.  In figure 1 we show the temperature
and abundance profiles obtained from the spectral fits, the values
reported for the innermost annulus are those obtained by leaving the
mass deposition rate as a free parameter.

By fitting the temperature and abundance profiles with a constant we
derive the following average values: $ 8.3\pm$0.3 keV and
0.35$\pm$0.03, solar units.  A constant fits adequately the
temperature profile; using the $\chi^2$ statistics we find, $\chi^2
=$3.1 for 4 d.o.f., corresponding to a probability (P) of 0.54 for the
observed distribution to be drawn from a constant parent distribution.
On the contrary, a constant does not provide an acceptable fit to the
abundance profile, $\chi^2 =$17.3 for 4 d.o.f.  (P $=$0.002).
Interestingly, a linear profile of the type, Ab = a $+$ b~r, where Ab
is in solar units and r in arcminutes, provides a better fit, $\chi^2
=$7.1 for 3 d.o.f. (P = 0.07). However, according to the F-test,
there is a relatively high probability (P = 0.15) that the
improvement may be associated to the reduction in the d.o.f. Indeed,
as clearly visible from figure 1, the abundance drop is seen only when
going from the first to the second bin, all other bins have abundances
consistent with the one derived in the second bin.
   
We have used the Fe K$_{\alpha}$ line as an independent 
estimator of the ICM temperature. 
We recall that the centroid of the observed Fe K$_{\alpha}$ line,
which is essentially a blend of the He-like Fe line at 6.7 keV, 
and the H-like Fe line at 7.0 keV, can be used to derive an estimate
of the gas temperature, a detailed description of the method we employ 
can be found in M99. 
Considering the limited number of counts available in the observed line we 
have performed the analysis on 2 annuli with bounding radii
0$^{\prime}$-6$^{\prime}$ and 6$^{\prime}$-12$^{\prime}$.
In figure 1 we have overlaid the temperatures derived from the 
centroid analysis on those previously obtained through the 
thermal continuum fitting. The two measurements of the temperature 
profile are clearly in agreement with each other.

\begin{figure}
\vspace{-0.16in}
\centerline{
      \hbox{
      \psfig{figure=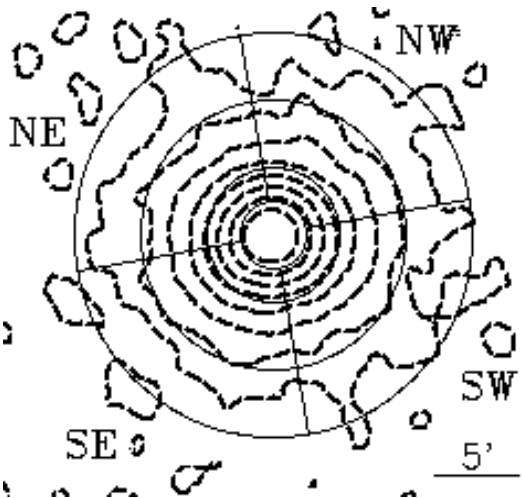,height=6.5cm,width=7cm,angle=0}
}}
\vspace{-0.16in}
\caption{
MECS image of PKS0745. Logarithmic contour levels are indicated
by the dashed lines. The thin lines show how the cluster
has been divided to obtain temperature and abundance maps.
}
\label{figmg7}
\end{figure}
\vspace{-0.15in}
We have divided PKS0745 into 4 sectors: NW, SW, SE and NE.  Each
sector has been divided into 3 annuli with bounding radii,
2$^{\prime}$-4$^{\prime}$, 4$^{\prime}$-8$^{\prime}$ and
8$^{\prime}$-12$^{\prime}$. In figure 2 we show the MECS image with
the sectors overlaid.  A correction for the absorption caused by the
strongback supporting the detector window has been applied for the
sectors of the 8$^{\prime}$-12$^{\prime}$ annulus.  We have fitted
each spectrum with a MEKAL model absorbed by the galactic N$_H$.  For
the spectra accumulated in the outermost annulus, where the background
contributes about $60\%$ to the total spectrum, we have taken into
account a $10\%$ variation in the background normalization when
computing the confidence intervals for the temperature and the
abundance.
\begin{figure}
\vspace{-0.19in}
\centerline{
      \hbox{
      \psfig{figure=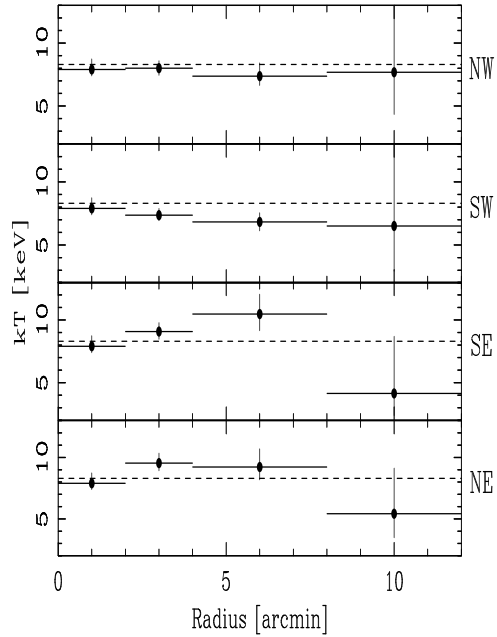,height=9cm,width=6.7cm,angle=-90}
}}
\vspace{-0.19in}
\caption{
Radial temperature profiles for the NW sector (first panel), the SW sector
(second panel), the SE sector (third panel) and the NE sector (forth
panel).
The temperature for the leftmost bin is derived from the entire circle,
rather than from each sector. 
The dashed lines indicate the average
temperature derived from the radial profile presented in figure 1.}
\label{figmg7}
\end{figure}

In figure 3 we show the temperature profiles obtained from the
spectral fits for each of the 4 sectors.  In all the profiles we have
included the temperature measure obtained for the central region with
radius 2$^{\prime}$ derived by the multiphase analysis detailed above.
Fitting each radial profile with a constant temperature we derive the
following average sector temperatures: 7.9$\pm$0.4 keV for the NW
sector, 7.3$\pm$0.4 keV for the SW sector, 8.7$\pm$0.5 keV for the SE
sector and 8.7$\pm$0.5 keV for the NE sector.  All sector averaged
temperatures are found to be within 2$\sigma$ of the average
temperature for PKS0745 derived from the radial profile reported in
figure 1.  From the $\chi^2$ statistics we find: $\chi^2 =0.3$ for 3
d.o.f. (P $= 0.96$) for the NW sector, $\chi^2 =1.0$ for 3
d.o.f. (P $= 0.80$) for the SW sector, $\chi^2 =3.5$ for 3
d.o.f. (P $= 0.32$) for the SE sector and $\chi^2 =2.9$ for 3
d.o.f. (P $= 0.41$) for the NE sector.  The temperature profile is
consistent with being constant in all sectors.  In the SE and NE
sectors the temperature of the second and third annuli is somewhat
larger than the temperature of the corresponding annuli in the NW and
SW sectors.  A fit with a constant to the temperatures of the second
and third annulus yields respectively: $\chi^2 =6.9$ for 3
d.o.f. (P $= 0.07$) and $\chi^2 =5.7$ for 3 d.o.f. (P $= 0.13$),
implying that the evidence in favor of an azimuthal gradient is not
particularly strong.

\begin{figure}
\vspace{-0.19in}
\centerline{
      \hbox{
       \psfig{figure=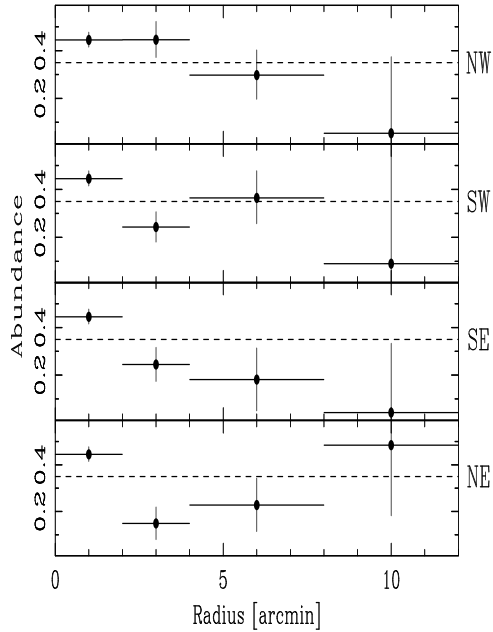,height=9cm,width=6.7cm,angle=-90}
}}
\vspace{-0.19in}
\caption{
Radial abundance profiles for the NW sector (first panel), the SW
sector (second panel), the SE sector (third panel) and the NE sector
(forth panel).  The abundance for the leftmost bin is derived from the
entire circle, rather than from each sector.  The dashed lines
indicate the average abundance derived from the radial profile
presented in figure 1.  }
\label{figmg7}
\end{figure}

In figure 4 we show the abundance profiles for each of the 4 sectors.
In all profiles we have included the abundance measure obtained for
the central region with bounding radius 2$^{\prime}$.  Fitting each
profile with a constant abundance we derive the following sector
averaged abundances: 0.43$\pm$0.03 for the NW sector, 0.40$\pm$0.03
for the SW sector, 0.40$\pm$0.03 for the SE sector and 0.38$\pm$0.03
for the NE sector.  The fits yield the following $\chi^2$ values:
$\chi^2 =3.2$ for 3 d.o.f. (P $= 0.36$) for the NW sector, $\chi^2
=8.3$ for 3 d.o.f. (P $= 4.0\times 10^{-2}$) for the SW sector,
$\chi^2 =10.9$ for 3 d.o.f. (P $=1.2\times 10^{-2} $) for the SE
sector and $\chi^2 =16.7$ for 3 d.o.f. (P $= 8.1\times 10^{-4}$)
for the NE sector.  A decreasing trend is observed in all sectors.  A
highly statistically significant gradient is observed only in the SE
and NE sectors.  Interestingly a drop in the abundance, when going
from the first to the second annulus, is observed for all sectors
except the NW. In the latter the abundance appears to decrease only
for radii greater than 6$^\prime$.  The abundance for the
2$^\prime$-4$^\prime$ annulus averaged over the SW, SE and NE sectors
is found to be different at more than the $99\%$ confidence level from
the abundance of the NW sector of the same annulus.

\section{Discussion}
The ASCA measurement of the average temperature of the ambient gas in
PKS0745 obtained by A96, $8.7^{+1.6}_{-1.2}$ keV (the errors are $90\%$
confidence), is in excellent agreement with the average temperature we
obtain from the BeppoSAX radial profile, $ 8.3\pm$0.3 keV.  Previous
measurements of the temperature structure of the ambient gas in
PKS0745 from ASCA data, by A96 and by White (1999), yield profiles
that are consistent with being constant. The rather large error bars
associated to the the measurements do not allow to place tight
constrains on the temperature profile.  The profile we report in this
Letter (see figure 1) is consistent with the one measured from
ASCA. Moreover, the better quality of the BeppoSAX data, allows us to
rule out, at more than the $99\%$ confidence level, a temperature
decrement of a factor 2 when going from the cluster core out to 1.2
Mpc.  The profiles we report in figure 3 suggest that the radial
temperature profile is constant for all sectors.  We find an
indication of an azimuthal temperature gradient occurring in the
annuli with bounding radii 2$^\prime$-4$^\prime$ (0.3 Mpc - 0.6 Mpc)
and 4$^\prime$-8$^\prime$ (0.6 Mpc - 1.2 Mpc).  The data suggests that
the SE and NE sectors of the cluster may be somewhat hotter than the
rest. Given the modest statistical significance of this temperature
gradient we refrain from pursuing the point any further.

No clear picture emerges from the radial abundance profile presented
by A96 (see their Fig 7b and their table 4), partially because of the
large errors and partially because of the discordance between the SIS
and GIS measurements. The abundance profile reported by White (1999)
is suggestive of an abundance gradient but statistically consistent
with a constant value. The radial abundance profile we report in this
Letter (see figure 1) shows a highly significant (more than $99.9\%$)
factor 2 drop in the abundance, when going from the first
0$^\prime$-2$^\prime$ (0-0.3 Mpc) to the second 2$^\prime$-4$^\prime$
(0.3-0.6 Mpc) radial bin.  From 0.3 Mpc out to 1.8 Mpc the profile is
consistent with being constant, note however that it is only out 0.9
Mpc that the errors allow us to exclude significant abundance
variations.  The profiles we report in figure 4 suggest that the
radial abundance gradient is most likely present in all sectors.
Interestingly the NW sector of the annulus with bounding radii
2$^\prime$-4$^\prime$ (0.3 Mpc - 0.6 Mpc) is found to have an
abundance similar to that derived for the core of the cluster and
significantly larger than the mean abundance derived from the other 3
sectors of the same annulus.  The presence of an abundance excess in
the NW sector with respect to the other sectors may indicate an
enhanced star formation rate. Intriguingly, the excess blue emission
associated to the central dominant galaxy discovered by Romanishin
(1987) and interpreted as the result of vigorous star formation is
found, by A96, to be marginally brighter in the NW direction.  Of
course the blue excess emission is seen on scales much smaller ($\sim$
10 kpc) than those on which we observe the abundance enhancement.
Nonetheless it is tempting to speculate that they both originate from
the same process, i.e. enhanced star formation in the NW direction.

\begin{acknowledgements}
We acknowledge support from the BeppoSAX Science Data Center.
\end{acknowledgements}


\end{document}